\documentclass[12pt]{iopart}

\usepackage[utf8]{inputenc}
\usepackage{graphicx}
\usepackage[caption=false]{subfig}

\usepackage{harvard}

\usepackage{amssymb}
\usepackage{amsopn}
\usepackage{bm}
\usepackage{stackrel}
\usepackage{iopams}
\usepackage{doi}

\DeclareMathDelimiter{\lvert}
  {\mathopen}{symbols}{"6A}{largesymbols}{"0C}
\DeclareMathDelimiter{\rvert}
  {\mathclose}{symbols}{"6A}{largesymbols}{"0C}
\DeclareMathDelimiter{\lVert}
  {\mathopen}{symbols}{"6B}{largesymbols}{"0D}
\DeclareMathDelimiter{\rVert}
  {\mathclose}{symbols}{"6B}{largesymbols}{"0D}
  
\usepackage[separate-uncertainty=true,list-units=single,range-units=single]{siunitx}
\usepackage{hyperref}

\providecommand{\keywords}[1]{\textbf{\textit{Index terms---}} #1}
\providecommand{\underset}[2]{\stackrel[#1]{}{#2}}

\begin{document}
\bibliographystyle{jphysicsB_withTitles}

\title[Quality assessment of MEG-to-MRI coregistrations]{Quality assessment of MEG-to-MRI coregistrations}
\author{Hermann~Sonntag$^{1,\,4}$, Jens~Haueisen$^{2,\,3}$ and Burkhard~Maess$^1$}
\address{$^1$ Max Planck Institute for Human Cognitive and Brain Sciences, Leipzig, Germany}
\address{$^2$ Institute of Biomedical Engineering and Informatics, Technische Universit\"at Ilmenau, Ilmenau, Germany}
\address{$^3$ Hans-Berger Department of Neurology, University Hospital Jena, Jena, Germany}
\address{$^4$ Author to whom any correspondence should be addressed.}

\ead{hsonntag@cbs.mpg.de}

\begin{abstract}
For high precision in source reconstruction of magnetoencephalography (MEG) or electroencephalography data, high accuracy of the coregistration of sources and sensors is mandatory.
Usually, the source space is derived from magnetic resonance imaging (MRI).
In most cases, however, no quality assessment is reported for sensor-to-MRI coregistrations.
If any, typically root mean squares (RMS) of point residuals are provided.
It has been shown, however, that RMS of residuals do not correlate with coregistration errors.
We suggest using target registration error (TRE) as criterion for the quality of sensor-to-MRI coregistrations.
TRE measures the effect of uncertainty in coregistrations at all points of interest.
In total, \num{5544} data sets with sensor-to-head and \num{128} head-to-MRI coregistrations, from a single MEG laboratory, were analyzed.
An adaptive Metropolis algorithm was used to estimate the optimal coregistration and to sample the coregistration parameters (rotation and translation).
We found an average TRE between \num{1.3} and \SI{2.3}{\milli\metre} at the head surface.
Further, we observed a mean absolute difference in coregistration parameters between the Metropolis and iterative closest point algorithm of \SI{1.9(15)}{\degree} and \SI{1.1(9)}{\milli\metre}.
A paired sample \textit{t}-test indicated a significant improvement in goal function minimization by using the Metropolis algorithm.
The sampled parameters allowed computation of TRE on the entire grid of the MRI volume.
Hence, we recommend the Metropolis algorithm for head-to-MRI coregistrations.
\end{abstract}

This is the Accepted Manuscript version of an article accepted for publication in \textit{Physics in Medicine and Biology}. IOP Publishing Ltd is not responsible for any errors or omissions in this version of the manuscript or any version derived from it. The Version of Record is available online at \doi{10.1088/1361-6560/aab248} and \doi{10.1088/1361-6560/aabf4e}.
This Accepted Manuscript is available for reuse under a CC BY-NC-ND 3.0 licence after the 12 month embargo period provided that all the terms of the licence are adhered to.
\maketitle

\section{Introduction}

The accuracy of the coregistration for magnetoencephalography (MEG) source reconstructions is limited by stochastic and systematic errors in the three measurement modalities involved: MEG, 3D-digitizer and magnetic resonance imaging (MRI).
While several suggestions have been made in the past to improve the accuracy of the coregistrations \cite{Singh1997,Adjamian2004,Troebinger2014a,Meyer2017}, no standard has been yet established.
In the present study we assess the quality of coregistrations using target registration error (TRE).
TRE is an error vector, of a point localization, resulting from coregistration uncertainties.
We propose a sequence of methods that are able to estimate TRE at any point of interest.

Coregistration procedures for MEG studies typically involve estimating sets of homologous positions, or coordinates, across at least two out of the three data modalities involved.
Each of the three modalities, (MEG, 3D-digitizer and MRI), provides a unique device coordinate system.
The MEG device coordinate system is defined by the MEG manufacturer to provide sensor positions.
MRI acquires an image relative to scanner-specific coordinates.
During 3D-digitization, anatomical landmarks are used to establish a subject-specific head coordinate system.
Within this paper, all positions will be reported relative to this head coordinate system.
The term \lq MEG coordinates\rq\ will refer to those which were originally given relative to the MEG device coordinate system and subsequently transformed to the 3D-digitized head coordinate system.
Likewise, coordinates which are extracted from an MRI scan and transformed to the 3D-digitized head coordinate system, will be referred to as \lq MRI coordinates\rq.
In practice, the results of brain activity studies are typically presented in head coordinates derived from brain internal fiducials only identifiable in structural MRI data, for example, MNI-coordinates\footnote{At the Montreal Neurological Institute (MNI), brain atlases were constructed from different sets of MR images. Different atlases are also named according to the number of MR images, which are the basis of the atlases (e.g.\ MNI305).} \cite{Evans1993}.

For convenience, we will use the following labels for the different coregistrations.
MEG to head coordinate transformations will be referred to as \textit{MEG-to-head} whereas head to MRI coordinate transformations will be referred to as \textit{head-to-MRI}. 
Both are assumed to be proper rigid transformations (rotation and translation). 
To assess the overall quality of the two coregistrations as a unit, they will be linked and referred to as \textit{MEG-to-MRI}.

There are a number of issues which contribute to coregistration uncertainty.
During MEG recordings the positions of the localization coils (coils for short) are estimated via magnetic field measurements and inverse modeling.
The solutions depend on signal quality and coil positions relative to the sensors \cite{Ahlfors1989,Fuchs1995}.
However, the coils make contact with the skin and can introduce error if their positions change while under tension.
Further, MRI scans may show systematic spatial deformations of the head shape, for instance due to air-filled cavities in the head or even via physical deformations of the head surface, for example by headphones.
In addition, estimation of the skin surface from MRI data depends on a threshold.
The extracted surface may therefore appear systematically above or below the actual skin surface.
According to \citeasnoun{Singh1997} defining anatomical landmarks, during the registration procedure, using two points on the ears and a third on the nasion only allows repeatability on the order of one millimeter at best.
The overall accuracy of the 3D-digitizer is influenced by the precision in digitizing the coil positions and the head shape.
However, during the digitization procedure these points can migrate slightly due to the elastic nature of the human skin.
Finally, coordinate transformations are based either on matching corresponding points (fiducials) between two coordinate systems or on surfaces (surface matching).
Pure fiducial based coregistrations are sensitive to fiducial localization errors and are highly likely to suffer larger errors than surface matching coregistrations when there are small numbers of fiducials \cite{Singh1997,Huppertz1998}. 


Several techniques have addressed the problem of fiducial localization errors.
One option is to fixate the participant's head using bite bars or head casts \cite{Singh1997,Meyer2017}.
Another common approach is to digitize the coils and head surface relative to an additional reference, attached to the subject's head \cite{Polhemus2012}.
This technique does account for head movements during digitization.
However, the methods proposed in this paper are also applicable to other \textit{MEG-to-head} coregistrations, which use either different definitions of the head coordinate system or additional mechanical means.

\citeasnoun{Schwartz1996} compared the two registration families (fiducial-based and surface matching) with respect to the \textit{head-to-MRI} coregistrations.
They used between \num{2000} and \num{4000} head shape points for surface matching and \num{3} points for pure fiducial-based registrations.
Their surface matching algorithm was based on a distance transform and the mean distance of all head shape points as cost function.
They reported an accuracy improvement for the surface matching technique compared to manual registrations.
The achieved accuracy of the registration was proportional to the number of head shape points.
Registration errors of \(\num{0.7} \pm \SI{0.3}{\milli\metre}\) were reported, estimated on a \SI{150}{\milli\metre} cube, sampled every \SI{2}{\milli\metre} using simulation tests. 
\citeasnoun{Huppertz1998} also estimated the accuracy of a surface matching technique for \textit{head-to-MRI} registrations for electroencephalography (EEG) data analysis. 
Between \num{1000} to \num{1800} head shape points were digitized and an iterative bisection search was used for surface matching.
They computed mean registration errors of \SIrange{1.4}{1.8}{\milli\metre} for \num{7} fiducial points using a test--retest design with \num{10} repetitions and \num{20} subjects.
The larger registration error compared to \citeasnoun{Schwartz1996} might be related to the points, where the registration error was measured.
More specifically, \citeasnoun{Schwartz1996} defined an equidistant grid in the MRI volume, while \citeasnoun{Huppertz1998} used \num{7} fiducials at the head surface.
Naturally, the points on the head surface show larger mean registration errors due to rotation uncertainties than fiducial points near the origin.
\citeasnoun{Wagner2001} used a similar approach to \citeasnoun{Huppertz1998} utilizing approximately \num{300} head shape points.
Their algorithm minimizes the \( L^1 \)-norm of the distances of head shape points to the MRI surface.
Unfortunately, no information about the achieved accuracy was provided.

There is substantial variability in the literature concerning \textit{head-to-MRI} coregistration methods.
For example, handheld laser scanners \cite{Koessler2011,Hironaga2014} and photogrammetry systems \cite{Koessler2007,Baysal2010,Qian2011} are proposed as alternatives to the electro-magnetic 3D digitization of electrode positions or head surface scanning.
\citeasnoun{Baysal2010} used a single camera photogrammetry system for EEG electrode localization and reported a maximum localization error of \SI{0.77}{\milli\metre} with 25 electrodes.
In a similar setting, \citeasnoun{Qian2011} reported a maximum localization error of \SI{1.19}{\milli\metre}.
They used 2 mirrors in addition to the system of \citeasnoun{Baysal2010}.
\citeasnoun{Koessler2007} compared a geodesic photogrammetry system with the \textit{Polhemus FASTRAK} and other electrode digitization techniques.
They reported an RMS position error of \SI{1.27}{\milli\metre} for the geodesic photogrammetry system and \SI{1.02}{\milli\metre} for the \textit{Polhemus}.
\citeasnoun{Koessler2011} tested EEG-to-MRI coregistrations using a 3D laser scanner.
An average of \num{5263} face shape points were recorded and an iterative closest point (ICP) algorithm was applied to the face shapes.
They reported a mean residual error of the electrode coregistration of \SI{2.11}{\milli\metre} for \num{65} electrode positions.
\citeasnoun{Hironaga2014} proposed a 3D laser scanner system for the \textit{MEG-to-MRI} coregistration.
They found superior registrations using the forehead surface compared to the upper head shape.
Further, they reported that TRE was at the submillimeter level using their regional registration method.
Our methods, proposed below, can be directly applied to data sets of the photogrammetry and laser scanner systems as mentioned above.

Previous studies have often only provided RMS of matched point residuals, for example, residuals of coil positions or head shape points, as a measure of the goodness of fit.
It has been shown, however, that these RMS of residuals and TRE are uncorrelated \cite{Fitzpatrick2009}.
Hence, the RMS of residuals are not well suited for determining the quality of the coregistrations.
Finally, previous studies concerned with the accuracy of coregistration measured or simulated TRE at only a few points \cite{Fuchs1995,Singh1997,Huppertz1998,Adjamian2004}.
In the present study we sample the distribution of coregistration parameters, and therefore TRE becomes a computable measure at any point of interest.
Consequently, we propose an overall assessment of the quality of individual coregistrations based on TRE.

\section{Methods}

\subsection{Instrumentation}

All data sets in our analysis were recorded using a \textit{Neuromag Vectorview} MEG with \num{102} planar magnetometers and \num{204} planar gradiometers.
In our laboratory, five localization coils are always used.
At the beginning of each measurement the five coils are energized by currents of unique frequencies.
This allows one to disentangle the superimposed fields and to estimate each coil's position, with respect to the MEG device, separately.
For the 3D-digitization of the coils and head shape, a \textit{Polhemus FASTRAK} system was used, which has a accuracy specification of \SI{0.8}{\milli\metre} RMS for all receiver positions in a radius of \SI{760}{\milli\metre} from the transmitter \cite{Polhemus2012}.
This distance is never exceeded in our lab.
The MRI surface extraction is based on the \textit{Freesurfer} segmentation of \SI{3}{\tesla} T1-weighted MPRAGE or MP2RAGE images with a voxel size of \SI{1x1x1}{\milli\metre}.

\subsection{Head coordinate system}

The definition of head coordinates depends on the MEG or EEG setup. 
In the present study \textit{Neuromag} head coordinates were used.
This coordinate system is often referred to as RAS, which is a mnemonic for the axes' pointing directions: right, anterior and superior.
The first axis of the head coordinate system is aligned with anatomical points on each ear, with coordinates increasing from left to right.
The second axis intersects perpendicularly, at the origin with the first, such that it runs through the nasion from posterior.
Thereby, the origin is not necessarily located at the middle between the ears.
Again, the third axis intersects at the origin, perpendicular to the first and second axes and coordinates are counted positive from inferior towards the subject's vertex.
This coordinate system was defined in \citeasnoun{Ahlfors1989} and is common for data acquisition with \textit{Neuromag} devices \cite[pages 25--26]{Neuromag2007}.

\subsection{Rotation by quaternions}

We used unit quaternions for the parametrization of rotations and their uncertainties for the following reasons.
Quaternions provide a convenient four-dimensional representation of object rotations.
They can be directly used to find the least squares solution of the coregistration of two corresponding point sets, while prohibiting reflections \cite{Besl1992}.
This is an advantage over the singular value decomposition based method, which permits reflections and may thereby yield an improper rotation matrix.
Furthermore, quaternion parameters provide an efficient method for three-dimensional rotations involving no trigonometric function computations.
The quaternion-based rotation is continuous over the unit sphere in \( \mathbb{R}^4 \).
The axis of a rotation is defined by a unit vector \( \vec{u} \).
A unit quaternion representing the rotation around \( \vec{u} \) by an angle of \( \theta \) is written as
\begin{eqnarray}\label{eq:def-quaternion}
\bm{q} &=& \exp{\left[\left( \theta / 2 \right) \left( u_1\bm{i} + u_2\bm{j} + u_3\bm{k} \right)\right]} \nonumber \\
&=& \cos{ \left( \theta / 2 \right) } + \left(u_1\bm{i} + u_2\bm{j} + u_3\bm{k}\right) \sin{ \left( \theta / 2 \right) } \nonumber \\
&=& q_0 + q_1\bm{i} + q_2\bm{j} + q_3\bm{k} \,,
\end{eqnarray}
where \(\bm{i}\), \(\bm{j}\) and \(\bm{k}\) represent the three imaginary units of quaternions.
Using \eref{eq:def-quaternion}, the rotation of a vector \(\vec{v}\) around \(\vec{u}\) by an angle of \(\theta\) is defined by 
\begin{equation}
\vec{v}' = \bm{q}\left( v_1\bm{i} + v_2\bm{j} + v_3\bm{k} \right)\bm{q}^{-1} = \bm{R}\left(\bm{q}\right)\vec{v}\,,
\end{equation}
where the inverse rotation quaternion \(\bm{q}^{-1}\) is simply obtained by converting the sign of the exponent in \eref{eq:def-quaternion} and \(\bm{R}\left(\bm{q}\right)\) denotes the respective rotation matrix as a function of \(\bm{q}\).
In the scope of this paper, the imaginary parts of the quaternion are referred to as rotation parameters and the real part is redundant for unit quaternions.
In order to evaluate rotations using a spatial distance, the rotation effect at a radius \( R \) is used.
On the plane orthogonal to the rotation axis, a rotation by an angle of \( \theta \) relates to a distance of \( R \cdot \theta \).
The relation of angles and unit quaternion parameters is derived from
\begin{equation}
    q_1^2 + q_2^2 + q_3^2 = \sin^2{ \left( \theta / 2 \right) }
\end{equation}
and for small angles \( \theta \approx 2\sqrt{ q_1^2 + q_2^2 + q_3^2} \). 
Hence the effect of rotations for points at the surface of a sphere, with a radius \( R \), is approximated by multiplying them (\( q_1\,, q_2\,, q_3 \)) with the diameter of sphere \( 2R \). This scaling is used in \sref{sec:sampling}, where the rotation parameters are sampled together with the translation parameters in the \num{6}-dimensional parameter space.
We selected \( R = \SI{100}{mm} \) as a scaling radius to approximate the radius of human heads.

\subsection{Coregistration model} \label{sec:model}

\subsubsection{MEG-to-Head}

This coregistration is based on \( M < 10 \) corresponding points, for example, coil positions.
Coil positions were first measured by the 3D digitizer and expressed in the head coordinate system. They are estimated in MEG device coordinates based on fitting a magnetic dipole field for each coil using \text{mne-python} \cite{Gramfort2013}.
The coregistration for the \textit{MEG-to-head} alignment of the points \(\bm{A} = \left(\vec{a}_1,\, \vec{a}_2,\, \dots,\, \vec{a}_M\right)\) localized in the \textit{MEG} and \(\bm{B} = \left(\vec{b}_1,\, \vec{b}_2,\, \dots,\, \vec{b}_M\right)\) digitized in the \textit{head} coordinate system is given by 
\begin{equation}
    \vec{b}_m = \bm{R}\left(\bm{p}\right)\vec{a}_m + \vec{s} + \vec{\epsilon}_m \,,\qquad m = 1,\, 2,\, \dots,\, M \,,
    \label{eq:meg2head}
\end{equation}
where the transformation is defined by the quaternion \(\bm{p}\) dependent rotation \(\bm{R}\) and the translation \(\vec{s}\) plus the error vector \(\vec{\epsilon}_m\).
The estimated solution to the coregistration problem is the set of parameters \(\hat{\bm{p}}\) and \(\hat{\vec{s}}\), which minimizes the residuals \(\vec{\delta}_m\) in the least squares sense according to
\begin{eqnarray}
	\hat{\bm{p}},\, \hat{\vec{s}} &=& \underset{\bm{p},\, \vec{s}}{\operatorname{argmin}} \sum_{m=1}^M \lvert \bm{R}\left(\bm{p}\right)\vec{a}_m + \vec{s} - \vec{b}_m \rvert^2 \label{eq:argmin_ls} \\
            \hat{\vec{b}}_m &=& \bm{R}\left(\hat{\bm{p}}\right)\vec{a}_m + \hat{\vec{s}} \\
			\vec{\delta}_m &=& \vec{b}_m - \hat{\vec{b}}_m \,.
\end{eqnarray}
We implemented the quaternion-based least squares solution for the problem in \eref{eq:argmin_ls} as proposed by \citeasnoun{Besl1992}.
For approximate parameter covariance estimation, the problem in \eref{eq:meg2head} is centred and linearized at the minimum of \eref{eq:argmin_ls} as
\begin{equation}
    \vec{b}^{\mathsf{c}}_m = \bm{J}_m \cdot \left( \tilde{p}_1, \tilde{p}_2, \tilde{p}_3, \tilde{s}_1, \tilde{s}_2, \tilde{s}_3 \right) + \vec{\epsilon}_m \,,\qquad m = 1,\, 2,\, \dots,\, M\,,
\end{equation}
where the superscript \(^{\mathsf{c}}\) denotes vector subtraction of the respective mean, \( \sum_{m=1}^M \vec{b}_m = \sum_{m=1}^M \hat{\vec{b}}_m \), and the Jacobians read
\begin{equation}
    \bm{J}_m = \left(
    \begin{array}{cccccc}
        0 & \hphantom{-}2\hat{b}^{\mathsf{c}}_{3m} & -2\hat{b}^{\mathsf{c}}_{2m} & 1 & 0 & 0 \\
        -2\hat{b}^{\mathsf{c}}_{3m} & 0 & \hphantom{-}2\hat{b}^{\mathsf{c}}_{1m} & 0 & 1 & 0 \\
        \hphantom{-}2\hat{b}^{\mathsf{c}}_{2m} & -2\hat{b}^{\mathsf{c}}_{1m} & 0 & 0 & 0 & 1
    \end{array} \right) \,,
\end{equation}
\cite{Wheeler1995}.
Under the assumption of homoscedastic errors \( \epsilon \) with zero mean and variance \( \sigma^2_{\epsilon} \), the parameter covariance matrix of the respective linear least squares estimate of the quaternion \( \tilde{\bm{p}} \) and translation \( \tilde{\vec{s}} \) yields
\begin{equation}
    \operatorname{Var}\left[ \tilde{p}_1, \tilde{p}_2, \tilde{p}_3, \tilde{s}_1, \tilde{s}_2, \tilde{s}_3 \right] = \sigma^2_{\epsilon} \cdot \left( \bm{J}^{\mathsf{T}}\bm{J} \right)^{-1} \label{eq:meg2head_covariance}
\end{equation}
\cite[equation (2.1.6)]{Bjoerck2015}, where \( \bm{J}^{\mathsf{T}} = \left( \bm{J}^{\mathsf{T}}_1,\,\bm{J}^{\mathsf{T}}_2,\,\dots,\,\bm{J}^{\mathsf{T}}_{M} \right) \).
As a result of the centring, there is no coupling between quaternion and translation parameters and two matrices are derived separately as
\begin{eqnarray}
     \operatorname{Var}\left[ \tilde{p}_1, \tilde{p}_2, \tilde{p}_3 \right] &=&  \sigma^2_{\epsilon} \cdot \left( 4 \sum^M_{m=1}{ \left( \lvert \hat{\vec{b}^{\mathsf{c}}}_m \rvert^2 \bm{I} - \hat{\vec{b}^{\mathsf{c}}}_m \hat{\vec{b}^{\mathsf{c}}}_m^{\mathsf{T}}\right)} \right)^{-1} \label{eq:quaternion_variance} \\
     \operatorname{Var}\left[ \tilde{s}_1, \tilde{s}_2, \tilde{s}_3 \right] &=& \sigma^2_{\epsilon} \cdot \bm{I} / M \label{eq:translation_variance}
    \,,
\end{eqnarray}
where \( \bm{I} \) is the identity matrix of size \num{3}.
The right hand expression of \eref{eq:quaternion_variance} is equivalent to a related variance estimate of \citeasnoun[equation (33)]{Markley2000}. 

\subsubsection{Head-to-MRI}

\begin{figure}[hb]
    \centering
    \includegraphics{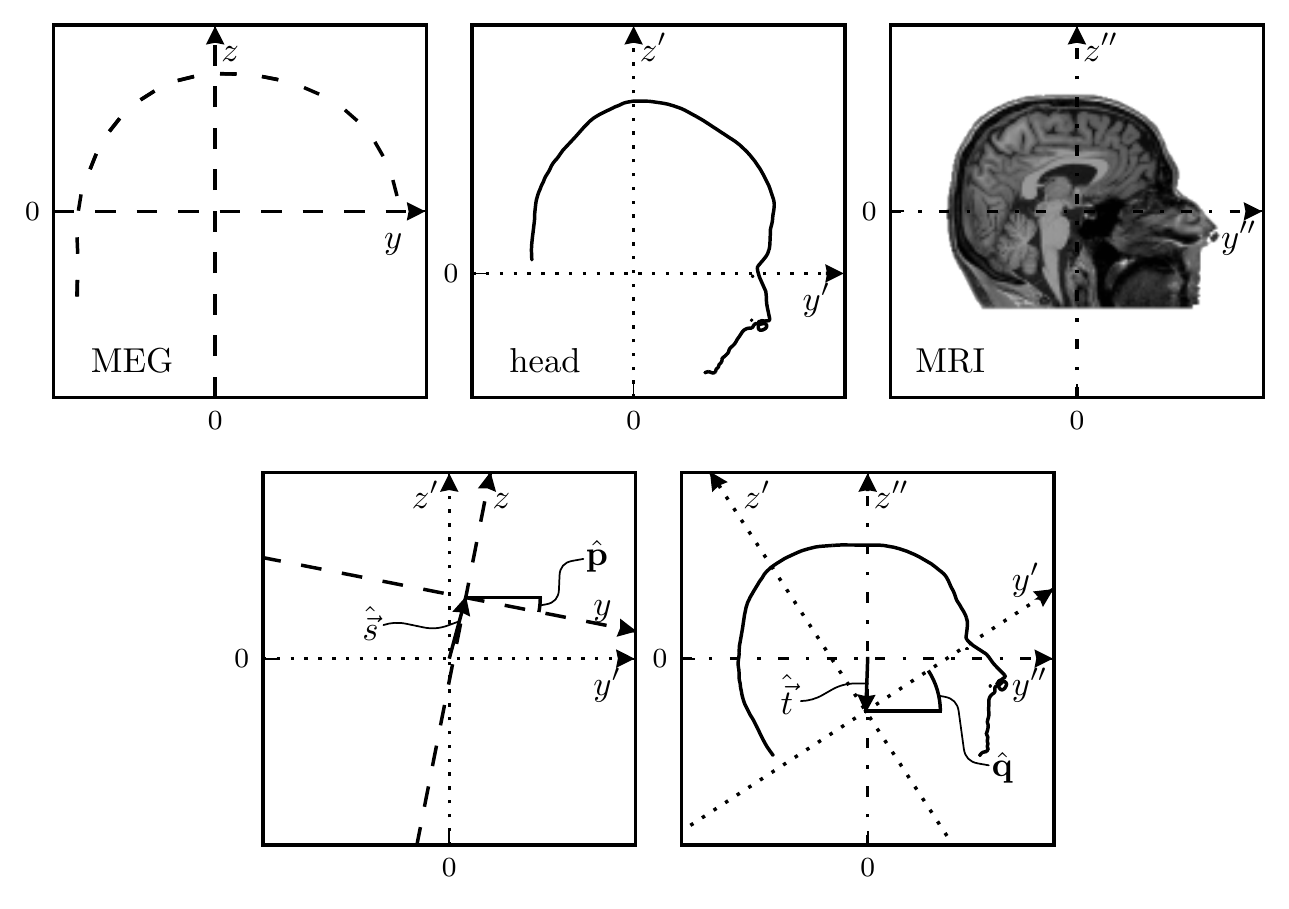}
    \caption{In the top row, the MEG, head and MRI coordinate systems are shown separately.
    The MEG coordinates are denoted by \( \left(y, z\right) \) and the respective axes are plotted by dashed lines relative to the contour of the MEG sensor configuration.
    Dotted lines represent the axes of the head coordinates \( \left(y', z'\right) \) and the head contour is outlined within the respective coordinate frame.
    The MRI coordinate axes are plotted by dash-dotted lines, the respective coordinates are denoted by \( \left(y'', z''\right) \) and a sagittal MRI slice is shown accordingly.
    In the bottom row, the notations and line styles are adopted from the top row and MEG/head and head/MRI coordinates are depicted relative to each other in the left and right box, respectively.
    The parameter notations \( \{ \hat{\bm{p}},\,\hat{\vec{s}} \} \) and \( \{ \hat{\bm{q}},\,\hat{\vec{t}} \} \) denote rotations and translations of \textit{MEG-to-head} and \textit{head-to-MRI}, respectively.
    Axes scaling is identical for all of the five sub-figures.
    }
    \label{fig:coordinates}
\end{figure}

This is a coregistration of \( N \sim 500\) points describing the head shape as measured by the 3D digitizer \( \bm{D} = \left(\vec{d}_1,\, \vec{d}_2,\, \dots,\, \vec{d}_N \right) \).
A second list with a point matrix \( \bm{E} \) is estimated via the segmented MRI data \( \bm{E} = \left\{ \vec{e}_1,\, \vec{e}_2,\, \dots,\, \vec{e}_P \right\} \). 
The subset \( \bm{F} = \left( \vec{f}_1,\, \vec{f}_2,\, \dots,\, \vec{f}_N \right) \) that best corresponds to \(\bm{D}\) depends on the quaternion \( \bm{q} \) and the translation \( \vec{t} \) and is the result of the closest point operator \( \mathcal{C} \), defined by
\begin{eqnarray}
        \vec{f}_n &=& \underset{\vec{f}}{\operatorname{argmin}} \lvert \bm{R}\left(\bm{q}\right)\vec{d}_n + \vec{t} - \vec{f} \rvert^2 \,, \vec{f} \in \bm{E} \\
        \bm{F} &=& \mathcal{C} \left(\bm{R}\left(\bm{q}\right)\bm{D} + \vec{t}\,\bm{1}^\intercal,\, \bm{E} \right) \,, \quad \bm{1}^\intercal = \left(1,\, \dots,\, 1\right) \in \mathbb{R}^{1\times N}
        \,.
    \label{eq:nearest_neighbor}
\end{eqnarray}
For the operator \( \mathcal{C} \), we used an efficient balltree implementation of the \textit{scikit-learn} module \cite{Pedregosa2012}.
Omitting the explicit notation of \( \mathcal{C} \), the \textit{head-to-MRI} problem reads as
\begin{equation}
	\vec{f}_n\left(\bm{q},\, \vec{t}\right) = \bm{R}\left(\bm{q}\right)\vec{d}_n + \vec{t} + \vec{\eta}_n \,,\qquad n = 1, 2, \dots, N
	\label{eq:head2mri}
\end{equation}
and a solution is
\begin{eqnarray}
		\label{eq:argmin_icp}
		\hat{\bm{q}},\, \hat{\vec{t}} &=& \underset{\bm{q},\, \vec{t}}{\operatorname{argmin}} \sum_{n=1}^N \lvert \bm{R}\left(\bm{q}\right)\vec{d}_n + \vec{t} - \vec{f}_n\left( \bm{q},\vec{t}\right) \rvert^2 \\
        \hat{\vec{f}}_n &=& \bm{R}\left(\hat{\bm{q}}\right)\vec{d}_n + \hat{\vec{t}} \\
		\vec{\zeta}_n &=& \vec{f}_n\left(\hat{\bm{q}},\, \hat{\vec{t}}\right) - \hat{\vec{f}}_n \,,
\end{eqnarray}
where \( \vec{\eta} \) and \( \vec{\zeta}_n \) are the error and residual vectors, respectively.
In realistic setups, the optimization problem of \eref{eq:argmin_icp} may not have a unique solution and due to the non-linearity of \( \mathcal{C} \), no closed-form solution is available.
Thus, an approximate solution is found using an iterative closest point (ICP) algorithm, which is likely to find local minima and therefore depends on the starting value \cite{Besl1992}.
Hence, the starting value was manually set by utilizing the 3D-digitized ear and nasion points and the 3D rendered MRI segmentation of the head shape.
The estimates \( \left\{ \hat{\bm{q}},\, \hat{\vec{t}} \right\} \) were computed by the ICP implementation in mne-python \cite{Gramfort2013}.
An overview of the coordinate system definitions and respective coregistration parameters is depicted in \fref{fig:coordinates}.

\newpage
\subsection{Data sets}

\subsubsection{MEG-to-Head} \label{sec:meg2head_datasets}

MEG data sets measured in our MEG laboratory in the years from 2007 to 2016 were retrieved from the archive and analysed.
For the present study, the term \lq MEG data set\rq\ denotes an MEG measurement block with coil position acquisition at the beginning of the block.
All included data sets had five coils attached to the participant's head.
We further restricted our selection to MEG data sets where none of the \num{204} gradiometers were marked as a bad channel.
In agreement with \citeasnoun[pages 38--39]{Neuromag2007}, two further quality control criteria were taken into account. 
First, the goodness-of-fit value for each coil had to be \num{0.98} or larger.
Second, the discrepancy between coil distances calculated from either MEG localization or from 3D digitization had to be smaller than \SI{5}{\milli\metre}.
In total, \num{7314} MEG data sets were considered, \num{5544} of them matched all of our selection criteria and formed the basis of the \textit{MEG-to-head} coregistration analysis.
A total of \num{1770} MEG data sets were rejected, \num{7} had bad gradiometers, \num{81} because of no coil measurement, \num{349} had less than \num{5} active coils, \num{405} because of the discrepancy between coil distances and \num{928} had goodness-of-fit values below \num{0.98}.

\subsubsection{Head-to-MRI}

For the \textit{head-to-MRI} coregistrations, only those MEG data sets were considered for which a segmented MRI data set was available and which included more than \num{200} head shape digitization points.
\textit{Head-to-MRI} coregistrations were conducted using \textit{MNE}, where the head surface extracted from MRI is matched with the 3D-digitized head shape using the ICP algorithm \cite[pages 195--197]{Hamalainen2010}.
Head shape points with a distance greater than \SI{10}{\milli\metre} from the MRI surface were excluded, as suggested by \citeasnoun[page 317]{Hamalainen2010}.
A total of \num{128} \textit{head-to-MRI} data sets were selected for the analysis.
A total of \num{149} \textit{head-to-MRI} data sets were rejected because they had less than \num{200} head shape points.
Most of the rejected data sets were from a time prior to our laboratory adopting more strict procedures. The recommended number of head shape points was increased over the years.

\subsection{Scales of the coordinate systems} \label{sec:scales_preliminary}

When coregistering data sets of different modalities, but from the same participant (i.e. the same head), one would not expect a need to scale the dimensions.
However, as briefly raised in the introduction, different methods may lead to systematic differences in the metrical scaling.
Thus far, we had assumed identical scalings in the different coordinate systems, that is, there is no change in length during the transformations.
This assumption, however, can be checked by analysing distance measures within each coordinate system separately.
The available data allowed pairwise comparisons of \textit{MEG} with \textit{head} and \textit{head} with \textit{MRI} coordinates.
To this end, we conducted a singular value decomposition (SVD) of the centred point clouds in both coordinate systems.
For convenience, we introduce the centring (demeaning) matrix for \( M \) points
\begin{equation}
    \bm{C}_{M} = \bm{I} - \frac{1}{M} \bm{1}\bm{1}^\intercal \,, 
\end{equation}
where \( \bm{I} \) is the identity matrix of size \( M \) and \( \bm{1}\bm{1}^\intercal \) is an \( M \times M \) matrix with each element equal to one.
For the centred point sets in the two coordinate systems \( \bm{A}^\mathsf{c} = \bm{A}\bm{C}_M \) and \( \bm{B}^\mathsf{c} = \bm{B}\bm{C}_M \), this reads as
\begin{eqnarray}
        \bm{A}^\mathsf{c} &=& \bm{U}_A \operatorname{diag}\left( \vec{\sigma}_A\right) \bm{V}_A ^\intercal \\
        \bm{B}^\mathsf{c} &=& \bm{U}_B \operatorname{diag}\left( \vec{\sigma}_B\right) \bm{V}_B ^\intercal \\
        c &=& \lvert \vec{\sigma}_A \rvert / \lvert \vec{\sigma}_B \rvert 
    \label{eq:scaling}
\end{eqnarray}
where \( \vec{\sigma}_A \) and \( \vec{\sigma}_B \) are the vectors of the positive singular values.
The scaling coefficient \( c \) between two systems is the quotient of the \( l^2 \)-norms of the singular value vectors.
\begin{table}[ht]
    \renewcommand{\arraystretch}{1.3}
    \caption{Scaling statistics of \textit{MEG-to-head} and \textit{head-to-MRI} are tested (two-tailed t-test).}
    \begin{indented}
        \item[]\begin{tabular}{@{}lSSS[table-figures-integer=2]S}
        \br
        \multicolumn{1}{l}{\textbf{Type}} & 
        \multicolumn{1}{l}{\textbf{mean}} & 
        \multicolumn{1}{l}{\textbf{SD}} & 
        \multicolumn{1}{l}{\textbf{\textit{t}-value}} & \multicolumn{1}{l}{\textbf{\textit{p}-value}} \\
        \mr
        \textit{MEG-to-head}  & 1.005 & 0.007 & 50.309 & \( < 0.001 \) 
 \\
        \textit{Head-to-MRI}  & 1.003 & 0.004 & 7.270 & \( < 0.001 \) 
 \\
        \br
    \end{tabular} 
    \label{tab:meg2head_scale}
    \end{indented}
\end{table}
\Tref{tab:meg2head_scale} shows a mean scaling of \( c \simeq 1.005 \) for \textit{MEG-to-head}, which translates to a \SI{0.5}{\milli\metre} difference at the head surface for a head radius of \SI{100}{\milli\metre}.
The expected error for the coil locations is in a similar range of about \SI{1}{\milli\metre} \cite{Ahlfors1989,Fuchs1995}.
Thus, we assume that the \textit{MEG} coordinates are systematically scaled by a factor of \num{1.005} and applied the correction to the \textit{MEG} coordinates.
The reason for this scaling effect might be the slight pressing force on the coils during digitization, which shifts the coils inwards and thus introduces a smaller scaling for digitization compared to MEG localization. 

\Tref{tab:meg2head_scale} shows a mean scaling of \( c \simeq 1.003 \) for \textit{head-to-MRI}, which results in a \SI{0.3}{\milli\metre} difference at the head surface.
Both scaling values were significantly different from \num{1}.
However, we have taken into account only the first and ignored the second.
This is because of the large variability between subjects at the level of the surface extraction from MRI data sets, in comparison to the estimated scaling value.
Furthermore, it is in agreement with \citeasnoun{Schwartz1996}, who state that surface matching is scaling independent if scaling effects are smaller than \SI{3}{\milli\metre}.

\subsection{Coil localization errors} \label{sec:coil_preliminary}

The \textit{MEG-to-head} coregistration is based on coil localizations.
\citeasnoun{Fuchs1995} investigated coil localization errors for three orthogonal coils (triplets), combined in a coil set, using a 31-channel Philips MEG.
They found that the coil localization error depends on the coil position relative to the sensor array as well as on the signal strength.
For a coil position below the sensor array they reported the difference between measured and true location to be less then \SI{1.8}{\milli\metre}, with a mean of \SI{1.1}{\milli\metre}.
The \textit{Neuromag Vectorview} device uses simpler single coils (no triplets) and it is a whole-head device with roughly ten times as many channels.
We investigated the device-specific error magnitude and its spatial dependency for data with \num{102} planar magnetometers and \num{204} planar gradiometers.
The coils were localized via their magnetic fields, each coil being modeled as a magnetic dipole \cite{Fuchs1995}.
Coil localization was exclusively based on the data of the 204 gradiometers because gradiometers have a higher signal to noise ratio for nearby sources due to their inbuilt suppression of distant (interfering) sources.
We estimated the variance of the noise via the norm of the misfit \( \bm{\chi} \) between the magnetic flux sensor signals \( \bm{s} \) and the modeled data 
\begin{eqnarray}
        \bm{\chi} \left( \hat{\vec{r}} \right) &=& \bm{s} - \bm{G} \left(\hat{\vec{r}}\right) \bm{G} \left(\hat{\vec{r}}\right)^{+} \bm{s} \\
        \sigma_{\mathsf{noise}}^2 &\sim& \frac{\lvert \bm{\chi} \rvert^2}{\num{204} - d} \,,
    \label{eq:noise}
\end{eqnarray}
where \( \bm{G} \left(\hat{\vec{r}}\right) \) is the leadfield of the magnetic dipole at \( \hat{\vec{r}} \) and \( \bm{G} \left(\hat{\vec{r}}\right)^{+} \) is the respective pseudoinverse.
The optimization has \( d = \num{6} \) degrees of freedom for each coil and we assumed that the noise follows an independent \textit{normal} distribution with zero mean, \( \sigma_{\mathsf{noise}}^2 \) variance and the respective probability density \( \pi_{\mathsf{noise}} \) in each channel.
Without prior knowledge about the parameters, the log-likelihood of the magnetic dipole location, given the measurement data, is defined by
\begin{equation}
    \log \pi \left( \vec{r} \mid \bm{s} \right) = \sum_{l=1}^{\num{204}} { \log \pi_{\mathsf{noise}} \left( \chi_l \left( \vec{r} \right) \right) } \,. 
    \label{eq:coil_likelihood}
\end{equation}
Samples are drawn from the probability density \( \pi \left( \vec{r} \mid \bm{s} \right) \) of the coil location, given the measurements, using the adaptive Metropolis algorithm of \citeasnoun{Haario2001} on the log-likelihood, see \eref{eq:coil_likelihood}.
We performed \num{10000} runs of the Metropolis algorithm, including \num{1000} burn-in samples.
In this test, \num{5x5544} coil positions of our \textit{MEG-to-head} data sets were included.
The maximal spatial error was only weakly dependent on the location in space. We estimated the dependency to \num{1.5e-3}, which represents \SI{0.15}{\milli\metre} at a distance of \SI{100}{\milli\metre}. Since this effect is about a \num{10}-th of the expected maximal error, we assumed equal coil localization errors for the volume of interest.
However, \citeasnoun{Fuchs1995} found a stronger dependency of the localization error on the position relative to the sensors.
This effect is likely related to the shape of the sensor array, as they used a 31-channel \textit{Phillips}-MEG with parallel sensor orientation and a smaller head coverage compared to the whole head, radially oriented sensor setup in the present study.

\subsection{Estimating errors from residuals} \label{sec:errors}

All residuals \( \vec{\delta}_m \) and \( \vec{\zeta}_n \), as defined in \sref{sec:model}, were separately concatenated from either \(K=\num{5544}\) \textit{MEG-to-head} or \(L=\num{128}\) \textit{head-to-MRI} coregistrations in the 
samples \( \Delta \) and \( Z \), respectively. 
The empirical distribution functions of a sample \( \Delta \) of size \( K \) is denoted by \( F_{\delta,\,K} \) and may be defined in terms of the order statistics \( \Delta_{(1)} \leq \Delta_{(2)} \leq \dots \leq \Delta_{(K)} \) by 
\begin{equation}
    F_{\delta,\,K} \left( x \right) = \cases{0 & if \(x < \Delta_{(1)}\)\\
    k / K & if \(\Delta_{(k)} < x \leq \Delta_{(k+1)},\, 1 \leq k < K\)\\
    1 & if \(x \geq \Delta_{(K)}\)\\}
\end{equation}
\cite[equation (2.1)]{Pratt1981}.
We modelled the distributions of the error elements of \( \vec{\epsilon}_m \) and \( \vec{\eta}_n \) using theoretical distributions for continuous random variables, for example, a logistic or a \textit{normal} distribution. However, the errors cannot be assessed directly. 
Therefore, the optimal theoretical distribution for the errors is chosen on the basis of the distributions of the residuals.
From a list of continuous candidate distributions we selected those with no, or one, shape parameter. These were implemented in \textit{scipy} and had good convergence (excluding \textit{rice} and \textit{erlang} distributions). Overall, these criteria resulted in a list of \num{69} distributions.
For the \( n \)-th candidate with distribution function \( G_{n,\,0} \left( x \mid \lambda_n,\, \mu_n,\, \sigma_n \right)\), the parameters shape \( \lambda_n \), mean \( \mu_n \) and scale \( \sigma_n \) were optimized according to
\begin{eqnarray}
        \hat{\bm{y}}_{\delta,\,n} &=& \underset{\lambda,\, \mu,\, \sigma}{\operatorname{argmin}} \left\{ \sup_x \bigg\lvert F_{\delta,\,K}\left( x \right) - G_{n,\,0} \left( x \mid \lambda,\, \mu,\, \sigma \right) \bigg\rvert \right\} \\
        \hat{\bm{y}}_{\zeta,\,n} &=& \underset{\lambda,\, \mu,\, \sigma}{\operatorname{argmin}} \left\{ \sup_x \bigg\lvert F_{\zeta,\,L}\left( x \right) - G_{n,\,0} \left( x \mid \lambda,\, \mu,\, \sigma \right) \bigg\rvert \right\} \\
        n &=& 1,\, 2,\, \dots \num{69}\,, \nonumber
    \label{eq:ks_one-sample}
\end{eqnarray}
where the optimization argument is the one-sample, two-sided Kolmogorov--Smirnov statistic \cite[equation (7.1)]{Pratt1981}.
The \textit{generalized normal} and the \textit{Students's t}-distribution yielded the smallest Kolmogorov--Smirnov statistics in \eref{eq:ks_one-sample} for the \textit{MEG-to-head} \( F_{\delta,\,K} \left( x \right) \) and \textit{head-to-MRI} \( F_{\zeta,\,L} \left( x \right) \), respectively.

The best fitting distributions were used as a basis to simulate residuals. 
Utilizing the \textit{generalized normal} distribution \( \mathcal{GN}\left(\lambda,\, 0,\, \sigma^2\right) \) for \( \tilde{\epsilon} \) we simulated \( \tilde{\delta}\left( \lambda,\, \sigma^2 \right) \) by replacing \(\vec{a}_m\) with \( \tilde{\vec{a}}_m = \vec{b}_m + \tilde{\vec{\epsilon}}_m \) in \eref{eq:argmin_ls}.
Accordingly, with the \textit{Student's t}-distribution with shape \( \lambda \) and scale \( \tau \) for \( \tilde{\zeta} \)
the residuals \( \tilde{\eta}\left( \lambda,\, \sigma^2 \right) \) are simulated by replacing \( \vec{d}_n \) with \( \tilde{\vec{d}}_n = \mathcal{C} \left( \hat{\vec{d}}_n,\, \bm{E} \right) + \tilde{\vec{\zeta}}_n \) in \eref{eq:argmin_icp}.
The two-sample, two-sided Kolmogorov--Smirnov statistics \cite[equation (3.1)]{Pratt1981}
\begin{eqnarray}
        D_{\tilde{\delta}} \left( \lambda,\, \sigma \right) &=& \max_x \bigg\lvert F_{\delta,\,K} \left( x \right) - F_{\tilde{\delta},\,K} \left( x \mid \lambda,\, \sigma \right) \bigg\rvert \,\,\, \text{and} \\
        D_{\tilde{\zeta}}\left(\lambda,\, \sigma\right) &=& \max_x \bigg\lvert F_{\zeta,\,L} \left( x \right) - F_{\tilde{\zeta},\,L} \left( x \mid \lambda,\, \sigma \right) \bigg\rvert
    \label{eq:ks_twosample}
\end{eqnarray}
were scanned for the set of parameters given in \tref{tab:meg2head_error_par}, which was selected in proximity of the optimum.
\begin{table}[ht]
    \renewcommand{\arraystretch}{1.3}
    \caption{Shape and scale parameters of the error distributions that were used to scan the Kolmogorov--Smirnov goal function.}
    \label{tab:meg2head_error_par}
    \begin{indented}
        \item[]\begin{tabular}{@{}llSS}
        \br
        \multicolumn{1}{l}{\textbf{Type}} & 
        \multicolumn{1}{l}{\textbf{name}} & 
        \multicolumn{1}{l}{\textbf{shape}} &
        \multicolumn{1}{l}{\textbf{scale in \si{\milli\metre}}} \\
        \mr
        \textit{MEG-to-head} & \textit{Gen. normal} & \( \num{1.7},\, \num{1.8} \dots \num{2.1} \) & \( \num{1.30},\, \num{1.35} \dots \num{1.55} \) \\
         & \textit{Normal} & & \( \num{0.90},\, \num{0.95} \dots \num{1.10} \) \\
        \hline
        \textit{Head-to-MRI} & \textit{Student's t} & \( \num{3},\, \num{4} \dots \num{7} \) & \( \num{0.90},\, \num{1.00} \dots \num{1.30} \) \\
        & \textit{Normal} & & \( \num{1.30},\, \num{1.35} \dots \num{1.60} \) \\
        \br
    \end{tabular}
    \end{indented}
\end{table}
Additionally, the \textit{normal} distribution was tested for comparison (\tref{tab:meg2head_error_par}).
Scanning of the Kolmogorov--Smirnov goal function is not deterministic since we drew samples from a distribution to simulate errors and residuals.
Therefore, error estimates of the Kolmogorov--Smirnov statistics were computed via multiple simulations of error distribution parameters, more specifically, \num{5} simulations for \textit{MEG-to-head} and \num{10} simulations for \textit{head-to-MRI}.
For \textit{head-to-MRI}, \num{5} simulations were insufficient because of higher variability in the corresponding Kolmogorov--Smirnov statistic.
The minimum of the Kolmogorov--Smirnov goal function corresponds to a certain distribution function,
which is taken as a model to approximate the error distribution.
Hence, these distribution parameters were utilized to sample the coregistration parameters in the following section.

\subsection{Coregistration parameter sampling} \label{sec:sampling}

In the previous section we approximated the distribution of errors for the point measurement in the coregistration problem of \eref{eq:meg2head} and \eref{eq:head2mri}.
We denoted the probability densities of the error distributions by \( \pi_{\epsilon} \) and \( \pi_{\eta} \) for \textit{MEG-to-head} and \textit{head-to-MRI}, respectively.
For the sampling of coregistration parameter distributions, we considered the centred and pre-registered problems.
The centring matrix transforms the coregistration points into their centred representation, for example, \( \bm{B}^\mathsf{c}\).
During pre-registration, coordinates from each modality are converted to head coordinates and aligned with the corresponding data set.
Having already applied a least squares or ICP optimization, all that remains in terms of error is the misalignment between the sets of data points and hence \( \hat{\bm{p}} = \hat{\bm{q}} = \bm{0} \) and \( \hat{\vec{s}} = \hat{\vec{t}} = \vec{0} \).

Log probability densities of a spatial error vector (e.g.\ \( \vec{a} \)) are defined by
\begin{equation*}
    \log \pi \left( \vec{a} \right) = \sum_{n=1}^3 {\log \pi \left( a_n \right)}\,.
\end{equation*}
The log-likelihood of the \textit{MEG-to-head} parameters \( \left\{\bm{p}, \vec{s}\right\} \), given the observation \( \bm{B}^\mathsf{c} \) and \( \hat{\bm{B}}^{\mathsf{c}} \) reads
\begin{equation}
    \log \rho \left( \bm{p},\, \vec{s} \mid \bm{B}^{\mathsf{c}},\, \hat{\bm{B}}^{\mathsf{c}} \right) = \sum_{m=1}^{M} \log \pi_{\hat{\epsilon}} \biggl[ \bm{R}\left(\bm{p}\right)\hat{\vec{b}^{\mathsf{c}}}_m + \vec{s} - \vec{b}^{\mathsf{c}}_m \biggr] \label{eq:meg2head_target}\,. 
\end{equation}
For the log-likelihood of the \textit{head-to-MRI} parameters \( \left\{\bm{q}, \vec{t}\right\} \), given the observation \( \bm{F}^{\mathsf{c}} \) and \( \hat{\bm{F}}^{\mathsf{c}} \), the additional closest point operator \( \mathcal{C} \) is required and \(\log \phi\) is therefore equivalently defined as
\begin{equation}
    \fl \log \phi \left( \bm{q},\, \vec{t} \mid \bm{E}^{\mathsf{c}},\, \hat{\bm{F}}^{\mathsf{c}} \right) = \sum_{n=1}^{N} \log \pi_{\hat{\eta}} \biggl[ \bm{R}\left(\bm{q}\right)\hat{\vec{f}^\mathsf{c}}_n + \vec{t} 
        - \mathcal{C} \left( \bm{R}\left(\bm{q}\right)\hat{\vec{f}^\mathsf{c}}_n + \vec{t},\, \bm{E}^{\mathsf{c}} \right) \biggr] \,. 
	\label{eq:head2mri_target}
\end{equation}
Utilising the log-likelihood, the target 
distributions of the parameters \( \bm{p},\, \vec{s}\) and \( \bm{q},\, \vec{t}\) given the observation, 
are sampled using a Metropolis algorithm 
on \eref{eq:meg2head_target}~and~\eref{eq:head2mri_target}, respectively.
Metropolis algorithms draw samples from an unknown distribution using samples from a known distribution, which is referred to as proposal distribution.
The original Metropolis algorithm uses a fix proposal distribution.
However, the convergence rate of the sample, to the desired unknown distribution, depends on the choice of the proposal distribution.
The adaptive Metropolis algorithm updates the proposal distribution by optimising the convergence using information from the sample chain at the current state.
\citeasnoun{Haario2001} used a Gaussian kernel proposal distribution with zero mean, hence only the proposal covariance needed updating.
An adaptive update scaling of the covariance of \( \num{2.4}^2 / d \) was used, following \citeasnoun{Haario2006}, with the dimensionality of the parameter-space \( d=\num{6} \).
The algorithm is non-Markovian but it has correct ergodic properties according to \citeasnoun{Haario2001}.
During parameter sampling, the adaptation of the Metropolis algorithm was performed for each step.
Before sampling, the rotation parameters were scaled by \( 2R = \SI{200}{\milli\metre} \) to homogenise the parameter space.
The initial proposal variance was set to \( \left( \SI{5}{\milli\metre} \right)^2 \) for the \textit{MEG-to-head} parameters and to \( \left( \SI{0.5}{\milli\metre} \right)^2 \) for the \textit{head-to-MRI} parameters based on prior experience.
We performed \( \num{e5} \) Metropolis algorithm iterations of the \textit{MEG-to-head} and \( \num{500} \times N \) iterations of the \textit{head-to-MRI} coregistrations, where \( N \) is the number of head shape points.
A burn in sample size of \num{1000} was used for both \textit{MEG-to-head} and \textit{head-to-MRI}.
The Metropolis sampling was implemented using the software library of \citeasnoun{muq}.
Since the adaptive Metropolis algorithm has correct ergodic properties, integral expressions over functions of the probability density of the parameters like the mean and the variance can be estimated by the respective expressions of sums over the functions on the sample.
Since the mean of the rotation parameters does not represent the mean rotation in general, we decided to provide the sample MLE instead of the mean.
In the expression of the variance of a parameter \( x \), the mean is replaced by the sample MLE accordingly as
\begin{eqnarray}
\int_{-\infty}^{\infty}{\rho \left( x \right) \left(x - x_{MLE}\right)^2 \mathrm{d}x} &\approx& \frac{1}{N}\sum_{n=1}^{N} \left(x_n - \hat{x}_{MLE} \right)^2 \nonumber \\
\operatorname{spread}\left( x \right) &=& \sqrt{\frac{1}{N}\sum_{n=1}^{N} \left(x_n - \hat{x}_{MLE} \right)^2}
\,,
\label{eq:spread}
\end{eqnarray}
where \( \rho \) is the probability density and \( N \) is the sample size.
Throughout this paper, the measure in \eref{eq:spread} is referred to as \lq spread\rq.

\subsection{MEG-to-MRI}

In the previous section, we referred to the centred and pre-registered problems for each of the two coregistrations (\textit{MEG-to-head} and \textit{head-to-MRI}) separately.
These centrings introduce a systematic shift between the translation parameters in the coordinate systems of both coregistrations.
However, taking this into account is straightforward.
One has to add the mean point  \( \bar{\vec{b}} \) of the first, and to subtract the mean point \( \bar{\vec{d}} \) of the second coregistration, that is, de-centring after the first and re-centring before the second transformation.
Consequently, the chained coregistration of a point \( \vec{a}^{\mathsf{\,MEG}} \) based on the \textit{MEG-to-head} and \textit{head-to-MRI}, as computed by the Metropolis algorithm, can be written as:
\begin{eqnarray}
        \vec{a}^{\mathsf{\,MRI}}_{k,\,l} &=& \bm{R} \left( \bm{q}_{l} \right) \cdot \biggl( \bm{R} \left( \bm{p}_{k} \right) \vec{a}^{\mathsf{\,MEG}} + \vec{s}_{k} + \bar{\vec{b}} - \bar{\vec{d}} \, \biggr) + \vec{t}_{l} \label{eq:meg2mri} \\
        \vec{a}_{\mathsf{MLE}}^\mathsf{\,MRI} &=& \bm{R} \left( \bm{q}_{\mathsf{MLE}} \right) \cdot \biggl( \bm{R} \left( \bm{p}_{\mathsf{MLE}} \right) \vec{a}^{\mathsf{\,MEG}} + \vec{s}_{\mathsf{MLE}} + \bar{\vec{b}} - \bar{\vec{d}} \,\biggr) + \vec{t}_{\mathsf{MLE}} \label{eq:mle} \,,
\end{eqnarray}
where \( \cdot_{\mathsf{MLE}}\) is the maximum likelihood estimate of the parameter from the Metropolis algorithm.
The indices \( k \) and \( l \) in \eref{eq:meg2mri} refer to the \( k \)-th and \( l \)-th subsample of \textit{MEG-to-head} and \textit{head-to-MRI} Metropolis samples, respectively.
For random sampling, \( k \) and \( l \) are drawn from the discrete uniform distribution of natural numbers between \( 1 \) and the corresponding Metropolis sample size.
Apart from the additional indexing, the notation is adopted from \eref{eq:meg2head} and \eref{eq:head2mri}, respectively (\fref{fig:coordinates}).
We defined TRE \( \vec{\psi} \) for the point \( \vec{a}^{\mathsf{\,MEG}} \) by
\begin{equation}
    \vec{\psi} \left( \vec{a}^{\mathsf{\,MEG}} \mid \bm{p}_{k},\, \bm{q}_{l},\, \vec{s}_{k},\, \vec{t}_{l} \right) = \vec{a}^{\mathsf{\,MRI}}_{k,\,l} - \vec{a}_{\mathsf{MLE}}^{\mathsf{\,MRI}} \,.
    \label{eq:tre}
\end{equation}
The RMS of TRE, defined by
\begin{equation}
    \operatorname{RMS} \left( \bm{\Psi} \right) = \sqrt { \frac{1}{G}  \sum_{g=1}^G { \lvert \vec{\psi}_g \rvert^2 }} \label{eq:rms_tre} \,,
\end{equation}
was used as a quality measure based on TRE at a specified point grid of size \( G \).
Statistics of \( \vec{\psi} \) and \( \operatorname{RMS} \left( \bm{\Psi} \right) \) were estimated by computation of \eref{eq:tre} and \eref{eq:rms_tre} for a large number of subsamples \( \left\{ \bm{p}_k,\,\vec{s}_k \right\} \) and \(  \left\{ \bm{q}_l,\,\vec{t}_l \right\} \). 

\section{Results}

\subsection{Errors and residuals}

\subsubsection{MEG-to-Head}

The smallest value for the maximal deviation measured by the Kolmogorov--Smirnov statistics (see \eref{eq:ks_twosample}), between the points and theoretical distributions was found for the \textit{generalized normal} distribution with shape \( \lambda = \num{1.7} \) and which estimated to \( D_{\tilde{\delta}} = \num{4.5e-3} \pm \num{0.6e-3} \).
The maximal Kolmogorov--Smirnov-value for the \textit{normal} distribution with scale \( \sigma_{\hat{\epsilon}} = \SI{1.05}{\milli\metre} \) was only slightly larger: \( D_{\hat{\delta}} = \num{6.2e-3} \pm \num{0.8e-3} \).
The \textit{normal} distribution is the special case of the \textit{generalized normal} distribution with shape \( \lambda = \num{2} \).
Hence, we decided to approximate the error distribution of \( \epsilon \) using the commonly used \textit{normal} distribution.
The probability density of the error estimate \( \hat{\epsilon} \) was therefore defined as
\begin{equation}
    \pi_{\hat{\epsilon}} \left( x \right) = \frac{1}{\sigma_{\hat{\epsilon}}\sqrt{2\pi }}\exp{\left[-{\frac{x^{2}}{2\sigma_{\hat{\epsilon}}^{2}}}\right]}\,.
\end{equation}
This choice provided control over our approximations, since closed form solutions are available under the precondition of the \textit{normal} distribution for the relation between variances (error, residual, and parameter) in a least squares estimation \cite{Fitzpatrick2009}.
The ratio between the variances of errors and residuals was found to be \( \sigma_{\hat{\epsilon}}^2 / \sigma^2_{\delta} = \num{1.65} \approx 5 / 3 \), which is approximately the ratio of the number of data points and the number of data points minus the degrees of freedom of the least squares fit, namely \( 3M / \left(3M - 6 \right) = M / \left(M - 2 \right) \).
\begin{figure}[b]
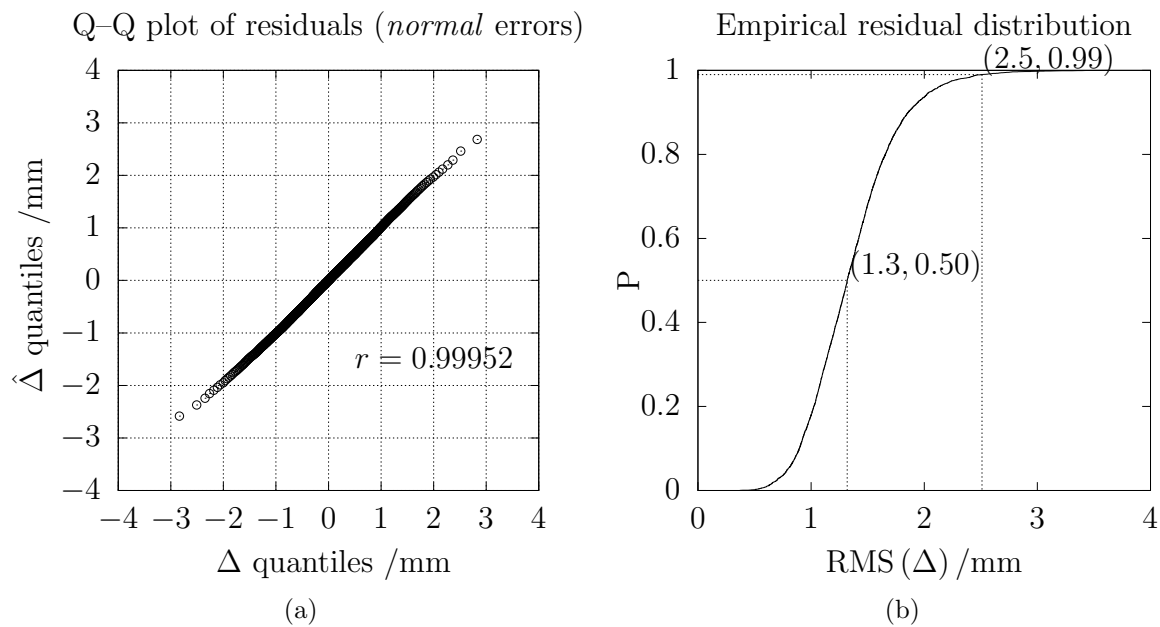

\subfloat[\label{fig:meg2head_qq}]{\input{fig2a}}
\subfloat[\label{fig:meg2head_rms}]{\input{fig2b}}
     \caption{The \textit{MEG-to-head} residuals Q--Q plot (a) depicts every \num{100}th data point of the \(\hat{\Delta}\)-quantiles over the \(\Delta\)-quantiles. The \( r \)-value is the correlation coefficient between the paired sample quantiles.
     The empirical distribution function of RMS of observed \textit{MEG-to-head} residuals is depicted in (b).
     }
\end{figure}
\Fref{fig:meg2head_qq} demonstrates the distribution-wise similarity between \(\hat{\Delta}\) and \(\Delta\) using a Q--Q plot, where \(\hat{\epsilon} \sim \mathcal{N}\left(0, \, \left(\SI{1.05}{\milli\metre}\right)^2\right)\).
If both distributions were identical, the Q--Q plot would show a straight diagonal.
Divergence from linearity at both ends show that the deviations between the two distributions were mainly observed with respect to the tails.
The residuals \(\hat{\Delta}\) and \(\Delta\) were distributed between \SIrange{-3}{3}{\milli\metre}, with approximately zero median and mean.
In \fref{fig:meg2head_rms}, the distribution of observed RMS of residuals is plotted for the \num{5544} \textit{MEG-to-head} data sets.
One RMS value is calculated over the \num{5} residual vectors \( \vec{\delta}_m \) of the coil positions.
\Fref{fig:meg2head_rms} shows that RMS values were smaller or equal to \SI{2.5}{\milli\metre} for \SI{99}{\percent} of the \textit{MEG-to-head} data sets.
The RMS values were distributed between \SIrange{0.4}{3.6}{\milli\metre}, with a median of \SI{1.3}{\milli\metre}.

\subsubsection{Head-to-MRI}

\begin{figure}
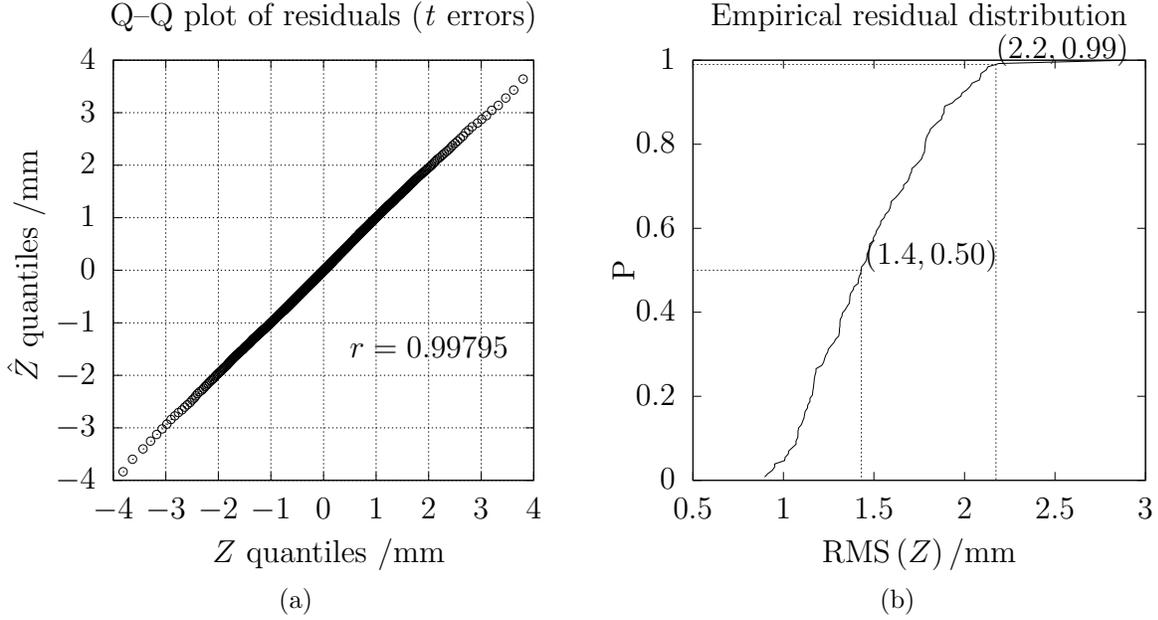

    \subfloat[\label{fig:head2mri_qq}]{\input{fig3a}}
    \subfloat[\label{fig:head2mri_rms}]{\input{fig3b}}
    \caption{The \textit{Head-to-MRI} residuals Q--Q plot (a) depicts every 100th data point of the \(\hat{Z}\)-quantiles over the \(Z\)-quantiles, where \( \hat{\eta} \) follows the \textit{t}-distribution with shape 4 and scale \SI{1.1}{\milli\metre}.
    The \( r \)-value is the correlation coefficient between the paired sample quantiles.
    The empirical distribution function of RMS of observed \textit{head-to-MRI} residuals is depicted in (b).
    }
\end{figure}
The smallest \( D_{\tilde{\zeta}} \)  was found for a \textit{Student's t}-distribution with shape \( \lambda = \num{4} \) and scale \( \tau = \SI{1.1}{\milli\metre} \) with \( D_{\hat{\zeta}} = \num{5e-3} \pm \num{1e-3} \).
Hence, the probability density of the error estimate \( \hat{\eta}\) is expressed efficiently as
\begin{equation}
    \pi_{\hat{\eta}} \left( x \right) \propto  \left(1+\frac{x^2}{\tau^2 \lambda} \right)^{-\left(\lambda+1\right) / 2} \,,
\end{equation}
directly proportional to a normalization constant.
We found a ratio between the variances of errors and residuals of \( \sigma_{\hat{\eta}}^2 / \sigma_{\zeta}^2 = \left( \lambda \tau^2 / \left( \lambda - 2 \right) \right) / \sigma_{\zeta}^2 = \num{2.87} \). 
The Q--Q plot in \fref{fig:head2mri_qq} demonstrates the similarity between \(\hat{Z}\) and \(Z\) in distribution, where \(\hat{\eta} \) follows the \textit{t}-distribution with shape 4 and scale \SI{1.1}{\milli\metre}.
Residual values of \(\hat{Z}\) and \(Z\) were in the range of \SIrange{-4}{4}{\milli\metre}, as indicated in \fref{fig:head2mri_qq}, with approximately zero median and mean.
The best fit \textit{normal} error distribution yielded substantially worse \textit{head-to-MRI} residuals with a Kolmogorov--Smirnov statistic of \( D_{\hat{\zeta}} = \num{9e-3} \pm \num{7e-4} \).
In \fref{fig:head2mri_rms}, the distribution of observed RMS of residuals is plotted for the \num{128} \textit{head-to-MRI} data sets.
One RMS value is calculated over the head shape point residuals \( \vec{\zeta}_n \) for each data set.
An RMS of up to \SI{2.2}{\milli\metre} was not exceeded for \SI{99}{\percent} of the \textit{head-to-MRI} data sets.
The RMS values were between \SIrange{0.8}{2.9}{\milli\metre}, with a median of \SI{1.4}{\milli\metre}.

\subsection{Parameter-distribution sampling}

\subsubsection{MEG-to-Head and head-to-MRI}

The MLEs and spreads of the coregistration parameters from the Metropolis algorithm samples were averaged over the data sets in \tref{tab:parameter}.
The first row in \tref{tab:parameter} demonstrates accurate estimates of the Metropolis algorithm with no differences compared to the least squares estimates.
For the \textit{MEG-to-head} data sets we found sample spreads of the Metropolis algorithm results of \SIrange{0.6}{0.9}{\milli\metre} for the scaled quaternion parameters and \SI{0.5}{\milli\metre} for the translations.

The spreads of \textit{MEG-to-head} parameters in \tref{tab:parameter} are identical, up to the first decimal place, to the theoretical estimate of \eref{eq:quaternion_variance} and \eref{eq:translation_variance}:
\begin{eqnarray}
\fl \SI{200}{\milli\metre} \cdot \sigma_{\epsilon} \cdot \sqrt{\operatorname{diag} \left[ \left( 4 \sum^M_{m=1}{ \left( \lvert \hat{\vec{b}^{\mathsf{c}}}_m \rvert^2 \bm{I} - \hat{\vec{b}^{\mathsf{c}}}_m \hat{\vec{b}^{\mathsf{c}}}_m^{\mathsf{T}}\right)} \right)^{-1} \right] } &=&
\left( \begin{array}{l}
\num{0.8(1)} \\ 
\num{0.9(1)} \\ 
\num{0.6}
\end{array} \right) \si{\milli\metre} \nonumber \\
 \sigma_{\epsilon} / \sqrt{M} &=& \SI{0.5}{\milli\metre}\,, \nonumber
\end{eqnarray}
where \( \sigma_{\epsilon} = \SI{1.05}{\milli\metre} \) and \( M = \num{5} \).
The numbers on the right hand side of the equation refer to sample means and standard deviations over the \num{5544} data sets.
This comparison provides a quality check of the Metropolis algorithm.

The results of the sample spreads, of the \textit{head-to-MRI} coregistration parameters in \tref{tab:parameter}, are similar to the results of \textit{MEG-to-head}, with slightly larger values in the scaled quaternion part of \SIrange{0.6}{1.0}{\milli\metre} and smaller values in the translation part of \SIrange{0.2}{0.4}{\milli\metre}.
Contrarily, the sample MLEs of \textit{head-to-MRI} in \tref{tab:parameter} show deviations up to several millimeters.
This indicates considerable difference between the pre-registration of the ICP and the subsequent registration of the Metropolis algorithm.
We found a mean absolute difference of the ICP compared to the Metropolis algorithm results of \SI{1.9(15)}{\degree} in the rotations and \SI{1.1(9)}{\milli\metre} in translations.
\begin{table}[ht]
     \renewcommand{\arraystretch}{1.3}
     \caption{Statistics of the Metropolis algorithm parameter results in \si{\milli\metre}.}
     \label{tab:parameter}
     \centering
     \begin{small}
         \begin{tabular}{@{}lSSS[table-figures-integer=2]SS[table-figures-integer=2]S}
         \br
         \multicolumn{1}{l}{\textbf{\textit{MEG-to-head}}} & 
         \multicolumn{1}{l}{\(2R \cdot p_1 \)} & 
         \multicolumn{1}{l}{\(2R \cdot p_2 \)} & 
         \multicolumn{1}{l}{\(2R \cdot p_3 \)} & 
         \multicolumn{1}{l}{\(s_1 \)} & 
         \multicolumn{1}{l}{\(s_2 \)} & 
         \multicolumn{1}{l}{\(s_3 \)} \\
         \mr
         MLE  & 0.0 & 0.0 & 0.0 & 0.0 & 0.0 & 0.0 \\
         Spread  & 0.8(1) & 0.9(1) & 0.6 & 0.5 & 0.5 & 0.5 \\
         \br
         \multicolumn{1}{l}{\textbf{\textit{Head-to-MRI}}} & 
         \multicolumn{1}{l}{\(2R \cdot q_1 \)} & 
         \multicolumn{1}{l}{\(2R \cdot q_2 \)} & 
         \multicolumn{1}{l}{\(2R \cdot q_3 \)} & 
         \multicolumn{1}{l}{\(t_1 \)} & 
         \multicolumn{1}{l}{\(t_2 \)} & 
         \multicolumn{1}{l}{\(t_3 \)} \\
         \mr
         MLE  & 0.6(33) & 0.4(24) & -0.2(13) & 0.1(7) & -0.2(12) & 0.1(3) \\
         Spread  & 0.9(3) & 1.0(3) & 0.6(2) & 0.3(1) & 0.4(1) & 0.2(1) \\
         \br
     \end{tabular}
     \end{small}
\end{table}
The respective paired differences of RMS of residuals were tested.
According to the \textit{t}-statistic, RMS computed by the Metropolis MLE were significantly smaller than RMS computed by the ICP fit with \( t = \num{3.04} \) and two-sided \( p < \num{0.01} \).
However, the difference of the means was only in the order of \SI{0.02}{\milli\metre}.

In order to test the correlation of \( \operatorname{RMS}\left( \bm{\Psi} \right) \) and RMS of residuals, we computed these measures separately for our \textit{MEG-to-head} and \textit{head-to-MRI} data sets.
\( \operatorname{RMS}\left( \bm{\Psi} \right) \) and RMS of residuals were computed separately over coil positions of \textit{MEG-to-head} and head shape points of \textit{head-to-MRI}.
Correlation coefficients were determined accordingly over the \num{5544} and \num{128} data sets.
We found correlation coefficients of \num{0.017} and \num{-0.116} for \textit{MEG-to-head} and \textit{head-to-MRI}, respectively.

\subsubsection{MEG-to-MRI}

\begin{figure}
    \input{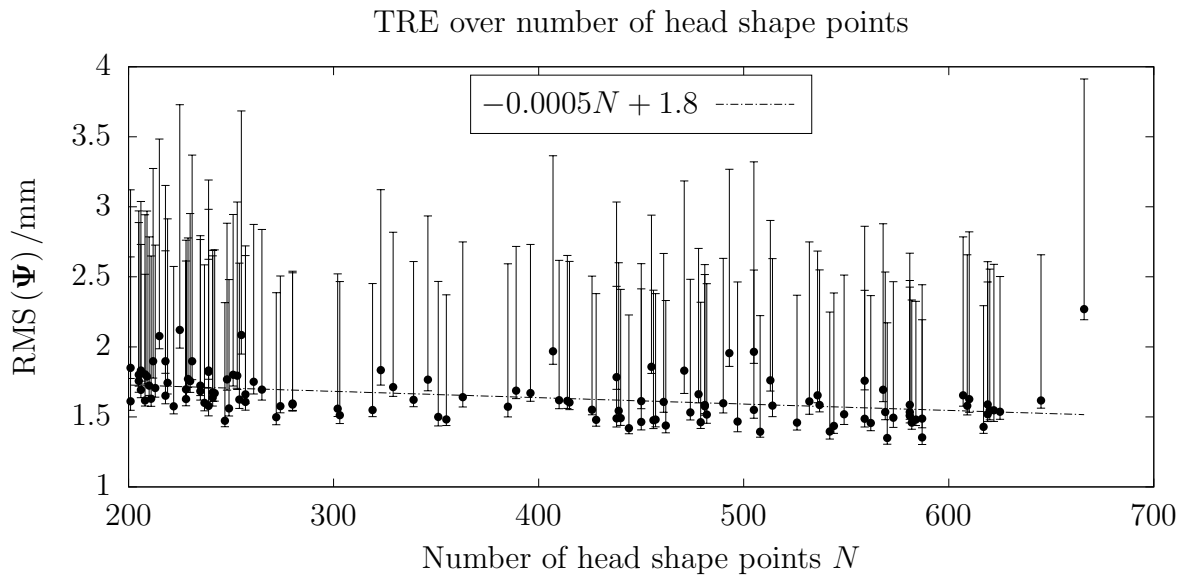}
    \caption{The estimated \( \operatorname{RMS} \left( \bm{\Psi} \right) \) is plotted over the number of head shape points N.
    \( \bm{\Psi} \) is computed at each head shape point.
    Data points indicate the mean over the samples of \( \operatorname{RMS} \left( \bm{\Psi} \right) \) and the dash-dotted line was fit to these points.
    The error bars show the \num{50}th to \num{95}th percentiles over the samples of the measure.}
    \label{fig:tre_nheadshape}
\end{figure}

\begin{figure}[t]
    \centering
    \includegraphics{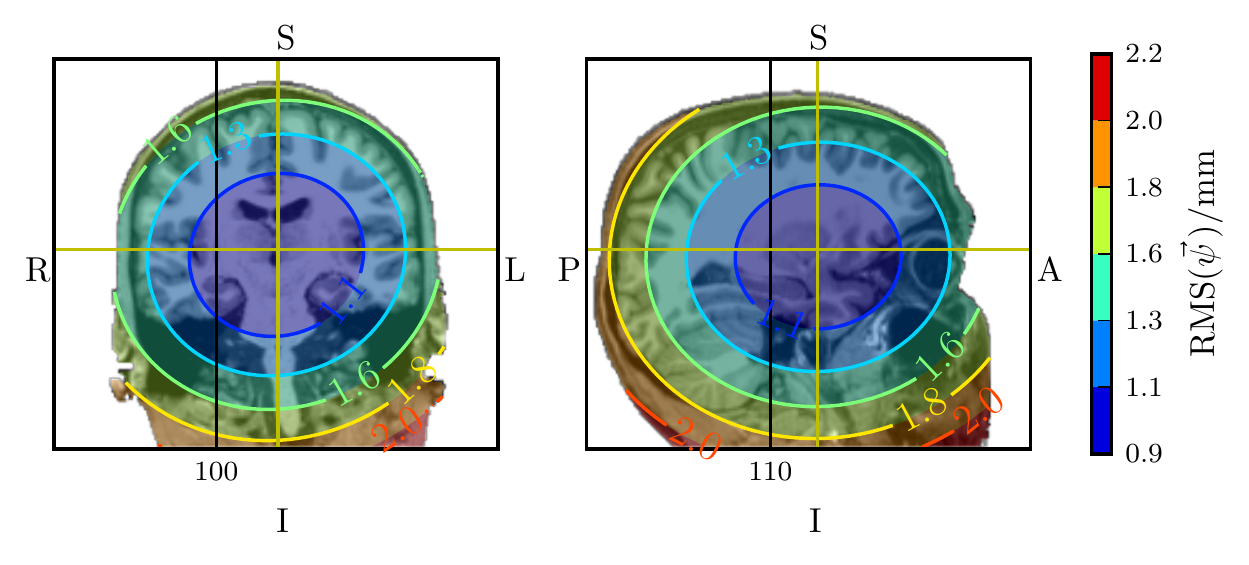}
    \caption{Estimates of TRE plotted as overlay onto the corresponding MRI slices.
    The RMS of TRE is computed for all samples of all grid points.
    Black lines indicate the slices in \textit{Freesurfer}-MRI coordinates.
    The yellow crosshairs indicate the estimated minimum of TRE.
    In the plots, A refers to anterior, P to posterior, I to inferior, S to superior, R to right and L to left.
    On the left and right side, the coronal and sagittal cuts at slice 110 and 100 are plotted, respectively.}
    \label{fig:tre_example}
\end{figure}

We found \num{126} out of the \num{128} \textit{head-to-MRI} data sets to have a corresponding \textit{MEG-to-head}, taking into account the selection criteria of \sref{sec:meg2head_datasets}.
If more than one \textit{MEG-to-head} data set corresponded to a given \textit{head-to-MRI}, which occurred if more than one MEG measurement block existed for a given session, only the first \textit{MEG-to-head} block was used.
\Fref{fig:tre_nheadshape} depicts the estimated RMS of TRE, denoted as \( \operatorname{RMS} \left( \bm{\Psi} \right) \), by the number of head shape points for these data sets.
The estimation of TRE is based on drawing subsamples from corresponding \textit{MEG-to-head} and \textit{head-to-MRI} Metropolis samples.
The size of the subsamples is the effective sample size of the respective Metropolis sample.
Utilizing these subsamples, the respective samples of the \( \operatorname{RMS} \left( \bm{\Psi} \right) \) were computed over the head shape points according to \eref{eq:tre}.
In a few cases there are multiple TRE data per head shape point numbers in \fref{fig:tre_nheadshape} due to coincidental digitization with the same number of points.
The error bars reflect the range, from the median to the \num{95}th percentile, over the samples of \( \operatorname{RMS} \left( \bm{\Psi} \right) \) whereas the points indicate the respective means.
We regard the \num{95}th percentile as an upper bound of the \( \operatorname{RMS} \left( \bm{\Psi} \right) \) confidence interval.
The data sets show a mean \( \operatorname{RMS} \left( \bm{\Psi} \right) \) of \SIrange{1.3}{2.3}{\milli\metre} and an upper bound of \SIrange{2.1}{4.0}{\milli\metre}.
Overall, both the mean and the upper bound decrease with the number of head shape points.
This TRE measure serves as a quality criterion for \textit{MEG-to-MRI} coregistrations and allows thresholding, for example, \SI{2}{\milli\metre}.
\Fref{fig:tre_example} shows the estimated TRE at a fine grid on the MRI of one data set.
Analogue to TRE, the coregistration rotation error is estimated by the RMS of \( \sqrt{q_1^2 + q_2^2 + q_3^2} \) for the \textit{MEG-to-MRI} rotation, which is easily sampled from the Metropolis algorithm results and does not depend on the position in space.
The angular approximation of this rotation error, estimated for each subject, is between \SIrange{0.8}{1.8}{\degree}, with the upper bound \num{95}th percentile between \SIrange{1.3}{3.1}{\degree}. 
The mean of the rotation error, across subjects, gives an angular approximation of \SI{1.1(2)}{\degree}.

\section{Discussion}
\subsection{Findings}
Using an adaptive Metropolis algorithm to sample the six-dimensional coregistration parameter space, and subsequent MLE, we were able to confirm the results of the least squares approach to \textit{MEG-to-head} coregistrations and further, to improve the results of the ICP algorithm for \textit{head-to-MRI} coregistrations.
As output, the Metropolis algorithm provides parameter sets with ergodic properties that allow confidence intervals of the coregistration parameters to be estimated.
Target registration error (TRE) is a function of the coregistration parameters, at any point in space, and statistical indices of TRE can be derived from the proposed Metropolis sampling.

We found that it is possible to approximate the empirical distributions of residuals in \textit{MEG-to-head} and \textit{head-to-MRI} coregistrations by replacing the point errors with samples from \textit{normal} and \textit{Student's t}-distributions respectively.
The empirical distributions indicated that \SI{99}{\percent} of the data sets yielded RMS of residuals of less than or equal to \SIlist{2.5;2.2}{\milli\metre} for \textit{MEG-to-head} and \textit{head-to-MRI} coregistrations respectively.
Thus, given our results, RMS values larger than these thresholds may indicate a problem in the measurement procedure.
However, this provides only a preliminary assessment where the given thresholds are exceeded in about \SI{1}{\percent} of the data sets.
Further, RMS of residuals are not well suited as a quality measure for coregistration, as they do not correlate with the actual errors (i.e. TRE) \cite{Fitzpatrick2009}.
This was confirmed in the present study where very small correlation coefficients, of \num{0.017} and \num{-0.116}, were observed over the \num{5544} \textit{MEG-to-head} and \num{128} \textit{head-to-MRI} data sets respectively.
For source reconstructions, TRE at the source location is the measure of interest.
TRE is the mislocalization of an alignment point due to uncertainty in the coregistration.
According to \eref{eq:tre}, we can estimate TRE distributions at any point in space if we can draw samples of the coregistration parameters.
An adaptive Metropolis algorithm can be used to sample the probability density of the coregistration parameters for each data set.
For the \textit{MEG-to-head} data sets, the MLEs of the Metropolis algorithm were equal to the least squares estimates.
This was the expected result as we used the probability density of a \textit{normal} distribution for the errors and in this case the least squares estimate is equal to the MLE \cite[equation 15.1.3]{Press1992}.
For the \textit{head-to-MRI} data sets, the Metropolis algorithm computed different MLE coregistration parameters compared to the ICP algorithm.
RMS of residuals were significantly reduced by the Metropolis algorithm compared to the ICP.
This may be explained by the fact that the ICP algorithm finds a local minimum dependent on the initial state of the iteration \cite{Besl1992}.
Coregistration optimizations like the \textit{head-to-MRI}, where only a subset of points in one modality correspond to the points of the other, depend on both the initial rotation and translation, and are also referred to as local shape-matching \cite{Besl1992}.
\citeasnoun{Besl1992} propose sampling the initial rotation and translation parameters for the local shape matching using the ICP algorithm.
However, this method is not common practice in MEG labs, and it is not implemented in \textit{MNE} or \textit{mne-python}, which are commonly used.
Compared to ICP, the Metropolis algorithm searches more globally and it is not completely determined by its initial state.
Samples can be drawn from the parameter distribution and variance, and higher moments can be estimated from the Metropolis samples because of the correct ergodic properties \cite{Haario2001}.
However, it should be noted that these advantages are achieved with higher computational costs compared to ICP.

For the translation parameter estimates, the \textit{head-to-MRI} yielded smaller variances compared to the \textit{MEG-to-head} coregistrations.
The high accuracy of the \textit{head-to-MRI} translation parameters can be explained by the larger number of data points compared to the \textit{MEG-to-head} coregistrations.
However, rotation parameters were similar between \textit{MEG-to-head} and \textit{head-to-MRI}.
This may be explained by the spherical nature of the head; spheres are rotation invariant in the \textit{head-to-MRI} coregistration problem.

For our data sets, we observed an RMS of TRE at the head surface of about \SI{1.7}{\milli\metre} on average.
We found an RMS of the rotation errors of about \SI{1.1}{\degree} on average, which was well predicted by the root of the sum over the squared quaternion spreads from \tref{tab:parameter}. 
\citeasnoun{Hillebrand2003} found a TRE threshold of \SI{2}{\milli\metre} at the cortical surface for anatomically constrained beamformers.
They suggest that the use of anatomical constraints with beamformers is only beneficial if the \textit{MEG-to-MRI} coregistration and segmentation error are smaller than \SI{2}{\milli\metre} and \SI{10}{\degree} at the cortex.
This result was later confirmed by \citeasnoun{Hillebrand2011} for the estimation of the source extent.
Our findings showed, on average, a smaller TRE than the critical \SI{2}{\milli\metre} value reported by \citeasnoun{Hillebrand2003} and \citeasnoun{Hillebrand2011}, but \SI{2}{\milli\metre} was still completely within the range of TRE distributions of the present study.
However, with respect to rotations, our results were consistently below the critical threshold of \SI{10}{\degree}.
We found an upper \num{95}th percentile of the coregistration rotation error of \SI{3.1}{\degree} at maximum.
The orientation of the cortical surface also depends on the segmentation, which may result in errors in the order of \SI{10}{\degree}.
In contrast to \citeasnoun{Hillebrand2003} and \citeasnoun{Hillebrand2011}, we did not assess TRE at the cortical surface but at head shape points because of the availability of this surface without conducting further segmentation.
However, using the Metropolis sampling of the coregistration parameters, we are able to compute TRE at any point in space.
For source reconstruction, TRE can be estimated at various points of interest in the source space or at the entire cortical surface.
For example, \fref{fig:tre_example} shows TRE computed on a coronal and sagittal slice.
The sagittal grid on the right side of the figure shows a small TRE in frontal regions of the brain.
These regions were close to the centre of the coil positions, where the \textit{MEG-to-head} produces the smallest TRE, and also close to the face, where the digitization provides more specific coregistration information compared to occipital regions.

Several studies have addressed the improvement in coregistration error stemming from particular measurement steps.
\citeasnoun{Singh1997} aimed to reduce the fiducial localization error effects using a bite bar.
They evaluated their strategy using Monte Carlo simulations and were able to substantially improve the stability of their coregistrations, in comparison to the pure fiducial-based method.
At the time of \citeasnoun{Singh1997}, tracking of head position and rotation, during head shape digitization, had not been established and, thus, the bite bar was essential to stabilise the head relative to the digitization reference.
A similar bite bar system was also proposed by \citeasnoun{Adjamian2004} which, reduced the fiducial localization error by approximately a factor of two.
They also reported that the bite bar can cause discomfort and introduces artifacts for some subjects.
In our laboratory, coils are placed freely on the anterior, upper part of the subject's head surface, independent of anatomical landmarks.
To compensate for head movement during 3D-digitization, head position and rotation are tracked using an additional reference, mounted on special glasses, which is common practice in present day MEG laboratories.
No additional mechanical hardware, for example, bite bars or individual head casts, are used to restrict the movement of the subject's head.
The methods of assessing coregistration errors suggested in the current report are not affected by mechanical hardware, although, if individual head casts are used a different approach for the assessment of the \textit{head-to-MRI} coregistration is needed.
\citeasnoun{Meyer2017} suggested the use of head casts that fit to the reconstructed surface of the MRI of individual subjects.
They estimated a maximal coregistration error of \SI{1.2}{\milli\metre} by using such head casts.
Depending on the shape of the subjects head, there was some flexibility in the positioning of the head, relative to the cast, which was tracked by a reference coil on the subject's nose, in addition to the coils in the cast.
They report a predominant uncertainty of about \SI{1.2}{\milli\metre} standard deviation of the head position relative to the cast in the z-axis (superiorly oriented head coordinate). 
However, potential movement of the subject's head, in a head cast, presents a problem that was not addressed by the assessments of our study.

Besides coregistration, head movement during data acquisition or between measurement blocks are related sources of error in MEG source reconstructions.
\citeasnoun{Uutela2001} compared two methods, a correction of sensor signals by alignment of minimum norm estimates and a correction of forward calculations.
They found that both methods can efficiently reduce the effect of head movement in typical MEG studies.
Later, an alternative method of sensor signal correction, based on multipole expansions, was proposed by \citeasnoun{Taulu2005} which is nowadays widely used with \textit{Neuromag} devices.
All of these methods rely on the accurate estimation of head positions during the MEG measurement.
Hence, their accuracy is intrinsically limited by the error of \textit{MEG-to-head} coregistrations.
The magnitude of head movements is often greater than the errors of \textit{MEG-to-head} coregistrations especially between measurement blocks and in studies with children.
For example, \citeasnoun{Wehner2008} reported an average head position displacement of \SI{12}{\milli\metre} from the beginning to the end of the experiment.
Compared to other sources of error, such as sensor noise and head movement, the \textit{MEG-to-MRI} coregistration error provides an absolute limit to the accuracy of source localization, which, cannot be reduced by longer measurements or sophisticated head movement corrections.

\subsection{Practical recommendations}

To facilitate a straightforward implementation of the proposed Metropolis algorithm for \textit{head-to-MRI} coregistration in different laboratories, we recommend the estimation of error variance from the residuals according to the ratio \( \sigma_{\hat{\eta}}^2 / \sigma_{\zeta}^2 = \num{2.87} \approx \num{3} \), which was found in the present study.
For the acquisition of \( \sigma_{\zeta}^2 \), we suggest the use of existing procedures from the respective laboratories (e.g.\ the ICP).
From this starting point, the estimation of error variance can be validated by error simulations and subsequent head shape matchings.
We recommend starting with variations of \textit{normal} or \textit{Student's t}-distributions.
As soon as a theoretical error distribution is found, with satisfying Kolmogorov--Smirnov statistics and a satisfying Q--Q plot of simulated and observed residuals, Metropolis sampling of the log-likelihood \eref{eq:head2mri_target} can be started.
For \textit{MEG-to-head} coregistration the Metropolis algorithm is not required in the case of approximately \textit{normally} distributed errors, of similar size as reported in the present study.
Assuming the latter conditions are met, parameter samples of \textit{MEG-to-head} can be generated by using \( \sigma^2_{\epsilon} \cdot \left( \bm{J}^{\mathsf{T}}\bm{J} \right)^{-1} \) of \eref{eq:meg2head_covariance} as the covariance matrix and a standard \textit{normal} random number generator.
The \textit{MEG-to-head} error variance \( \sigma^2_{\epsilon} \) can be estimated from residuals as \( \sigma^2_{\epsilon} = \sigma^2_{\delta} M / \left( M - 2 \right) \).
This is the theoretical ratio for linear least squares fits \cite[page 214]{Bjoerck2015} of rotation and translation parameters, where \( M \) is the number of coils.
Optimal coregistration parameters are found in closed form for \textit{MEG-to-head} and from the maximum likelihood estimate of the Metropolis sample for \textit{head-to-MRI}.
For corresponding parameter samples of \textit{MEG-to-head} and \textit{head-to-MRI}, TRE is estimated by computation of \eref{eq:meg2mri}, \eref{eq:mle} and \eref{eq:tre}.

Concerning the \textit{head-to-MRI} data sets of our lab, the emphasis of facial features, (e.g.\ bridge of nose) was used along with a large number of head shape points.
Hence, it is difficult to determine the exact contributions, to TRE, of the sheer number of points involved and the number of facial features used.
Taking into account the spatial  distribution of TRE, in \fref{fig:tre_example}, we suggest it might also be beneficial to acquire more head shape points in areas with the highest errors, such as the inion, which tends to have unique spatial  features.
A similar argument can be made for coil placement.
The hair complicates the attachment of the coils at occipital regions, which is the reason for a more frontal coil placement in our laboratory.
If possible, we recommend attaching at least one coil to an occipital location.
We recommend using a large number of head shape points, about \num{600} yielded the smallest TRE in the current study, emphasis on facial features as well as the inion.
However, the sheer number of head shape points is not a guarantee for good coregistration.
As seen in \fref{fig:tre_nheadshape}, the largest number of head shape points resulted, accidentally, in the largest TRE.
Therefore, and in agreement with \citeasnoun{Hillebrand2003} and \citeasnoun{Hillebrand2011}, we recommend checking that the mean RMS of TRE is not greater than \SI{2}{\milli\metre} at the head surface.

Computations of TRE, like in \fref{fig:tre_example}, are useful for coil placement and head shape digitization optimizations in EEG applications as well.
For example, for accurate reconstructions of brain activity in the visual cortex it is beneficial to refine the head shape digitization at occipital regions.
In this case, TRE at the visual cortex is the measure of interest.
Coregistrations for EEG only involve the \textit{head-to-MRI} problem although head shape digitization is more challenging due to the electrode cap, compared to the MEG procedure.
As a result of the electrode cap, the number of head shape points is usually smaller in the EEG coregistration compared to the equivalent procedure in MEG.
For this reason, the uncertainties of the fit are likely to be higher for EEG compared to the results of the present study.
We believe that the availability of TRE at regions of interest would be useful for the digitization optimization in EEG.

\section{Conclusion}
Quality assessment of \textit{MEG-to-MRI} coregistrations can be achieved by using the Metropolis sampling algorithm of the coregistration parameters and subsequently evaluating TRE.
Further, we propose establishing this assessment procedure in EEG and MEG laboratories and suggest reporting TRE in the study publications, especially if source estimates are reported.
We recommend the application of the Metropolis algorithm to achieve higher accuracy when estimating the parameters of the \textit{head-to-MRI} problem. 
Due to the superior results compared to the ICP, and the availability of parameter distribution samples and derived measures like TRE, we suggest the Metropolis algorithm also for EEG coregistration fits.

\section*{Acknowledgements}
The authors would like to thank Shameem Wagner and Joshua Grant for valuable comments on the preparation of the manuscript.

\section*{References}
\bibliography{bibliography}
\end{document}